# A Semiconductor Under Insulator Technology in Indium Phosphide

K. Mnaymneh,[1,2,3] D. Dalacu,[2] S. Frédérick,[2] J. Lapointe,[2] P. J. Poole,[2] and R. L. Williams[2,3]

[1]Department of Electrical and Computer Engineering, University of Michigan, Ann Arbor, MI, 48109 USA

[2]National Research Council Canada, 1200 Montreal Road, K1A 0R6, Ottawa, Canada

[3]Department of Physics, University of Ottawa, Ottawa, K1N 6N5, Canada

This Letter introduces a Semiconductor-Under-Insulator (SUI) technology in InP for designing strip waveguides that interface InP photonic crystal membrane structures. Strip waveguides in InP-SUI are supported *under* an atomic layer deposited insulator layer in contrast to strip waveguides in silicon supported *on* insulator. We show a substantial improvement in optical transmission when using InP-SUI strip waveguides interfaced with localized photonic crystal membrane structures when compared with extended photonic crystal waveguide membranes. Furthermore, SUI makes available various fiber-coupling techniques used in SOI, such as sub-micron coupling, for planar membrane III-V systems.

Photonic crystal (PhC) membranes in III-V materials offer versatile photonic dispersion engineering in an optically active medium. In addition to being an extremely attractive route for designing out-of-plane laser and optical sensing devices[1,2], III-V PhC membranes are also a natural choice for controlling the quantum electrodynamics of epitaxial confinement systems. For example, site-specific growth techniques can be used to deterministically place single quantum dots in photonic crystal nanocavities[3] achieving high single photon extraction ratios into adjacent PhC membrane waveguides[4,5]. Because conventional strip waveguides have lower loss than PhC membrane waveguides, it would be ideal to adiabatically transfer photons from the PhC membrane waveguides to these lower-loss strip waveguides. However, unlike strip waveguides in SOI where the high dielectric contrast confines the optical mode completely in the top silicon layer, it is not possible to make high dielectric contrast strip waveguides in III-V systems. For example, an InP strip waveguide supported on an InGaAs layer does not have the dielectric contrast to support a mode completely confined in the top InP layer that interfaces PhC membranes. Efforts have been made to circumvent this limitation. For example, one could use in-plane tether structures along the length of the released InP strip waveguide[6] as a support mechanism or hybrid integration using adhesion bonding of SOI strip waveguides with top III-V thin layers for active control of the confined optical modes[7]. However, in-plane tether structures lead to increased scattering at the tether contact points limiting the transmission bandwidth[6] and hybrid integration of SOI waveguides with III-V layers is a highly complex process leading to packaging issues and lower device yield[8]. An innovative way around this problem is to make an inverted version of the SOI system. Using an insulator layer on top of the InP as the supporting structure and removing the InGaAs layer directly underneath the InP strip waveguide, one can design strip waveguides in III-V systems akin to SOI strip waveguides and exploit fiber coupling methods in III-V systems that are normally available to SOI systems.

This Letter presents a semiconductor-under-insulator (SUI) technology for membrane compound semiconductors where the compound semiconductor used in this work is InP. We fabricate InP strip waveguides supported *under* an insulator layer in contrast to silicon strip waveguides supported *on* an insulator layer. Figure 1a shows a typical InP PhC-W1 waveguide terminated at the edge of the chip. It would be ideal to avoid this and keep the PhC-W1 membrane localized to a small area of the chip in order to minimize losses associated with fabrication and diffraction[9]. As in the case of SOI PhC-W1 membrane structures, losses are avoided by designing access waveguides that interface the membrane and run to the edge of the chip for fiber coupling. Figure 1b and 1c shows the facet edge of an InP strip waveguide supported from above by a conformal insulator layer of alumina that was deposited by atomic layer deposition (ALD). Figure 1c shows a close-up of the waveguide facet. The level of conformality by ALD provides excellent support for such access waveguides.

Starting with an InP substrate, a 1 μm of InGaAs was grown followed by 300 nm of InP. The InGaAs layer will serve as the sacrificial layer and the 300 nm layer of InP will be the top optical layer where the access waveguide and PhC membranes structures will be fabricated. A 90 nm of silicon oxide is then deposited by a low temperature plasma-enhanced chemical vapor deposition (PECVD) followed by 700 nm of electron beam resist. The silicon oxide layer serves as a hard mask for etching into the InP. Electron beam lithography is then used to pattern the strip waveguides and PhC membrane section at the same time. The PhC membrane parameters consisted of a lattice constant of 441 nm with a lattice hole diameter of 220 nm ensuring the air light-line is crossed at a wavelength around 1500 nm and having a maximum transmission at 1550 nm, as will be shown later in figure 3. The thickness of the PhC membrane is equal to the thickness of the top grown InP, which is 300 nm. After post-exposure development, the silicon oxide is dry etched using an induced-coupled plasma dry etcher. The electron beam resist is then removed before using the silicon oxide hard mask to dry etch the pattern into the top InP layer and a portion of the InGaAs layer. After removing the silicon hard mask, a 200 nm layer of alumina is then deposited using ALD. Release windows for wet etching the InGaAs sacrificial layer straddle the waveguides are then patterned and dry etched into the alumina. Before removing the InGaAs sacrificial layer, the alumina covering the PhC membrane is removed in order to have the common air/InP/air membrane structure. Photoresist is spin coated onto the sample with only the area above the membrane exposed. After development, the sample is dipped into hydrofluoric acid for 5 minutes to completely remove the alumina covering the membrane. The InGaAs sacrificial layer below the membrane is removed by dipping the sample in a 3:1:1 concentration of sulfuric acid, hydrogen peroxide and DI water, respectively for 30 seconds. The finished structure is shown in figure 1d with the inset showing the interface between the alumina held strip waveguide (slightly thicker portion of the access waveguide as pointed to in the inset) and the part of the waveguide going to the membrane.

The highly conformal nature of the ALD process is important to support the strip waveguides. Using a less conformal method to deposit the top insulator layer will result in air pockets leading to an inability to hold the waveguides. The conformality also makes available advanced fiber-coupling techniques normally available for SOI systems, such as submicron coupling and inverse tapers[10]. Having this capability in III-V membrane systems, where single quantum dots can be coupled to planar waveguides quite easily[5], is very important for increasing the optical collection efficiency into and from fiber optic networks for various next-generation applications.

High optical confinement in the InP-SUI strip waveguides is critical when creating mode-matching conditions for interfacing with highly confined optical modes of localized PhC membranes. With InP having a slightly lower refractive index than silicon[11] and alumina having a slight higher refractive index than silicon oxide[12], the dielectric contrast of the two systems should be very similar. However, the geometrical conditions are different: the silicon strip waveguide rests on the silica whilst the InP strip waveguide is conformally held from above by the alumina. An established commercially available software package[13] was used to study and compare the optical confinement in both types of strip waveguides. The calculation method is an eigenvalue finite-element analysis[14] of the following equation,

$$\nabla \times \nabla \times \mathbf{E}(x,y,z) - n^2(x,y)k_0^2 \mathbf{E}(x,y,z) = \mathbf{0}, \qquad (1)$$

where $\mathbf{E}(x,y,z)$ is the electric field vector of the optical signal in the waveguide. The form of the vector sought is, $\mathbf{a}_z E(x,y)e^{-i\beta z}$, where $E(x,y)$ is the electric field profile function perpendicular to the direction of propagation $z$ with a propagation constant $\beta$. The input parameters are the wavevector magnitude of $k_0$, with a free-space wavelength of 1550 nm, and the refractive index profile, given by $n(x,y)$. The eigenmode $E(x,y)$ and eigenvalue $\beta$ are calculated for both the SOI and InP-SUI strip waveguides, respectively and we report the eigenvalues here as the effective index, $n_{eff} = \beta/k_0$, for each waveguide. The calculation results, presented in figure 2a and 2b, show that the electric field eigenmode in both cases is very strongly confined to the silicon and InP waveguides with the calculated effective indices of the two types of strip waveguides being $n_{eff}^{SOI} = 2.56$ and $n_{eff}^{SUI} = 2.28$. Because the effective indices are similar and the confinement is high in both cases, guidance in InP-SUI strip waveguides is very similar to SOI waveguides implying techniques used for effective coupling in SOI should also be well suited for InP access waveguides.

Optical transmission of the InP-SUI strip waveguides, with and without an integrated PhC membrane, was obtained using a standard in-plane optical tabletop setup, as shown in the inset of figure 3. The samples were placed on a three-axis stage with the optical input coming from a butt-coupled fiber system consisting of a tunable laser source, polarization rotation optics and a microlensed fiber. The waveguide transmission output was collected by a microscope objective then sent through a polarizing beamsplitter before registering with a photodetector and displayed on a computer. Using the input polarization rotation optics, the input polarization was set to transverse electric (TE), with the electric field oriented in the x-direction indicated in the inset of figure 3, since the PhC membranes were designed to have their waveguide modes defined in their TE band gap. The wavelength range swept with the tunable laser was from 1300 nm to 1600 nm in order to fully characterize the optical passband of the PhC membranes. The optical transmission of two 1 mm long InP-SUI strip waveguides, (a) without and (b) with an integrated 20 μm long PhC-W1 membrane, was collected. The optical transmission of a (c) 500 μm end-to-end PhC-W1 membrane (whose edge facet is shown in figure 1a) was also collected and compared to the two InP-SUI strip waveguides (a) and (b). Figure 3 shows the results of the optical transmission through these three waveguides. The familiar rising and falling edges of PhC optical transmission are seen in (b) and (c) depicting the confined optical modes under the light-line and near the slow-light band-edge, respectively. The loss in the InP-SUI strip-membrane system, (b), is about 10 dB less than the end-to-end PhC-W1 membrane, (c), even though the end-to-end PhC-W1 membrane is half the length of the InP-SUI strip-membrane system. Such results clearly demonstrate that the top insulator supported InP strip waveguides is a solution for planar optical circuitry that desire to exploit the optical physics of PhC membranes in active media whilst collecting light in an efficient manner to and from fiber optic networks.

We have demonstrated a semiconductor-under-insulator technology for designing highly confined single mode strip waveguides in membrane compound semiconductor systems. This technology provides an effective way of fabricating strip waveguides in III-V systems facilitating active planar lightwave circuitry to take advantage of localized PhC membranes without incurring losses associated with long PhC membrane waveguides. The top insulator layer supporting the InP strip waveguide does not limit the transmission band as in the case of tethered supported strip InP waveguides and the active and passive optical material is InP unlike hybrid integrations involving SOI bonded to thin III-V layers. Furthermore, coupling techniques currently in use for SOI systems, such as submicron and inverse couplers, can easily be implemented in this system.

The authors gratefully acknowledge the help of Mr. Dan Roth, Mr. Mark Malloy and Ms. Martha E. McCrum.

FIG. 1. (a) The edge facet of an extended InP PhC-W1 membrane, (b) the edge facet of an InP strip waveguide supported under alumina deposited by ALD, (c) enlarged image of dashed box in (b) showing the InP strip waveguide supported by the conformal ALD alumina layer, (d) InP-SUI strip waveguide interfacing a PhC-W1 membrane structure. Inset is an enlarged image of the area in the dashed box showing the interface between the alumina-supported InP strip waveguide and the PhC-W1 membrane. The arrow in the inset points to a slightly thicker region of the strip waveguide identifying the alumina support layer.

FIG. 2. Finite element eigenvalue analysis of (a) silicon on insulator with an effective index of 2.56 and (b) InP under insulator with an effective index of 2.28. Calculation results show high confinement of the electric field for both SOI and InP-SUI waveguides. Mode propagation is in the z direction as indicated by the axis.

FIG. 3. Optical transmission with associated waveguide SEM image of (a) 1 mm InP-SUI strip waveguide, (b) 1 mm InP-SUI strip waveguide with a 20 μm integrated PhC W1 membrane in the middle of the waveguide (as shown in figure 1c) and (c) a 500 μm end-to-end PhC-W1 membrane (where edge facet of this waveguide is shown in figure 1a). Figure inset illustrates the transmission setup used to measure output transmission of the waveguides.

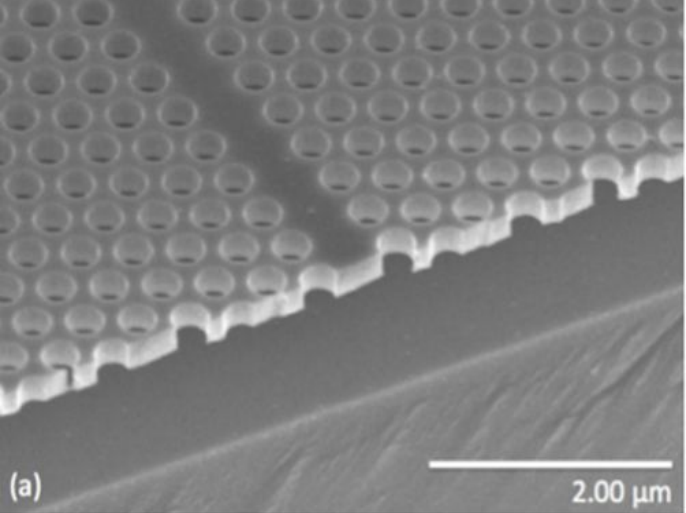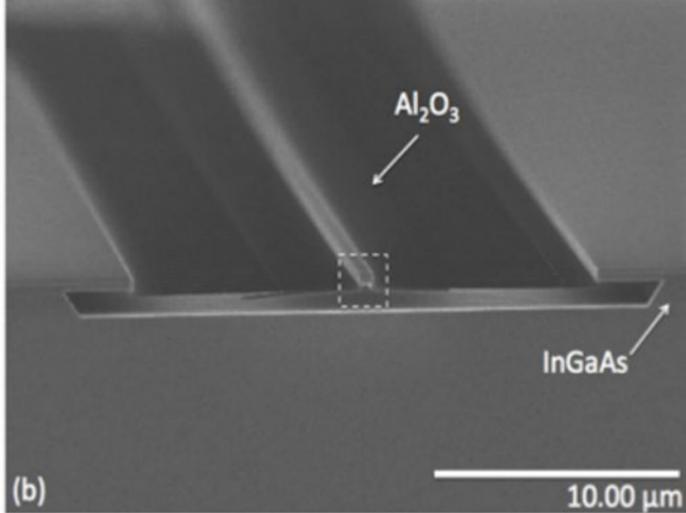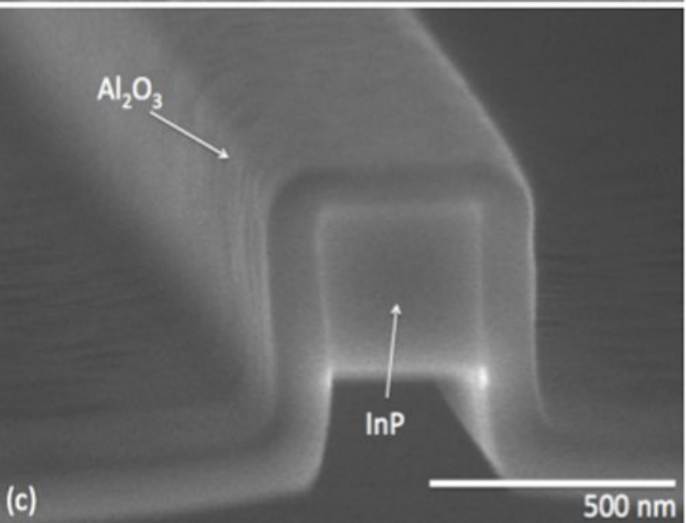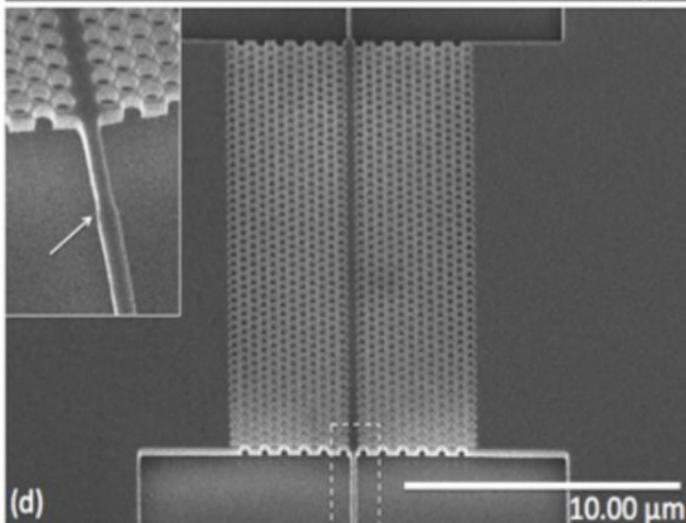

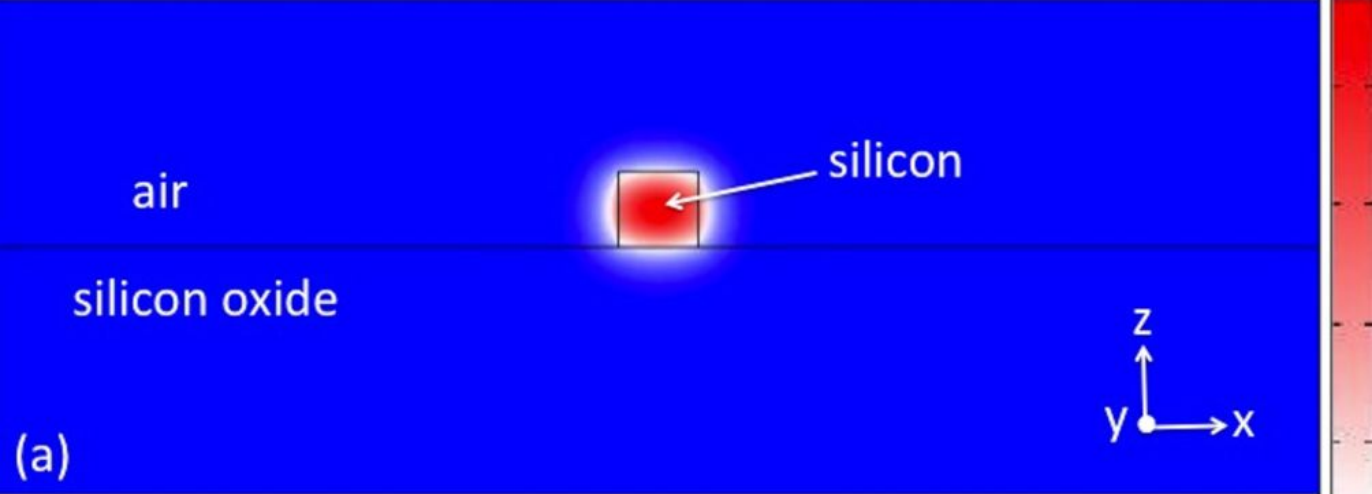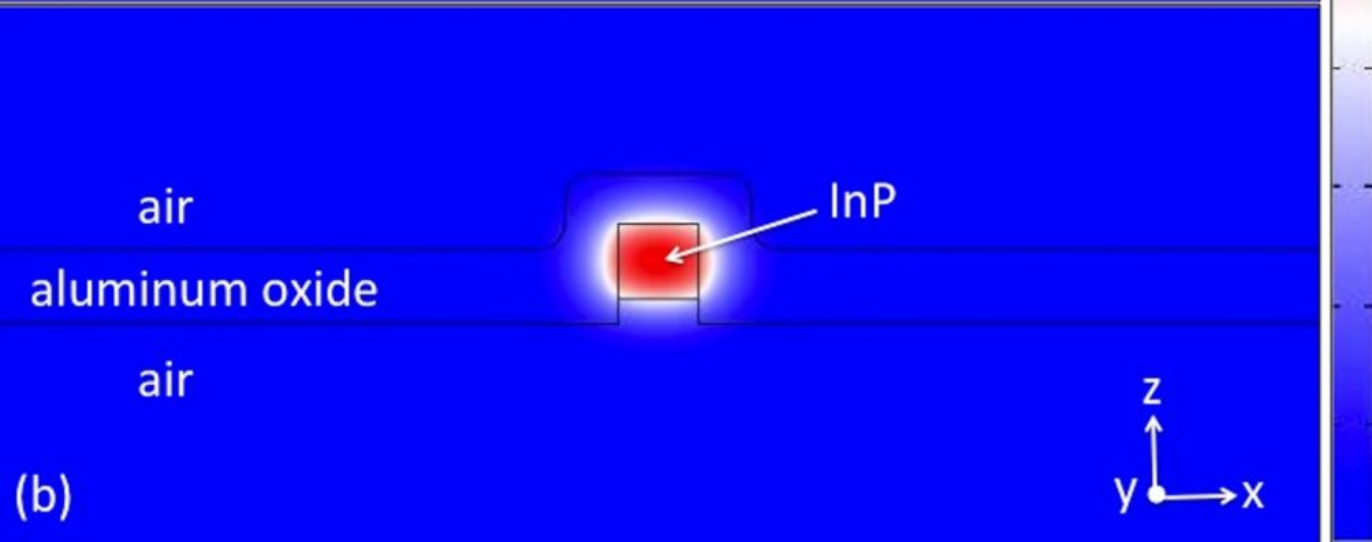

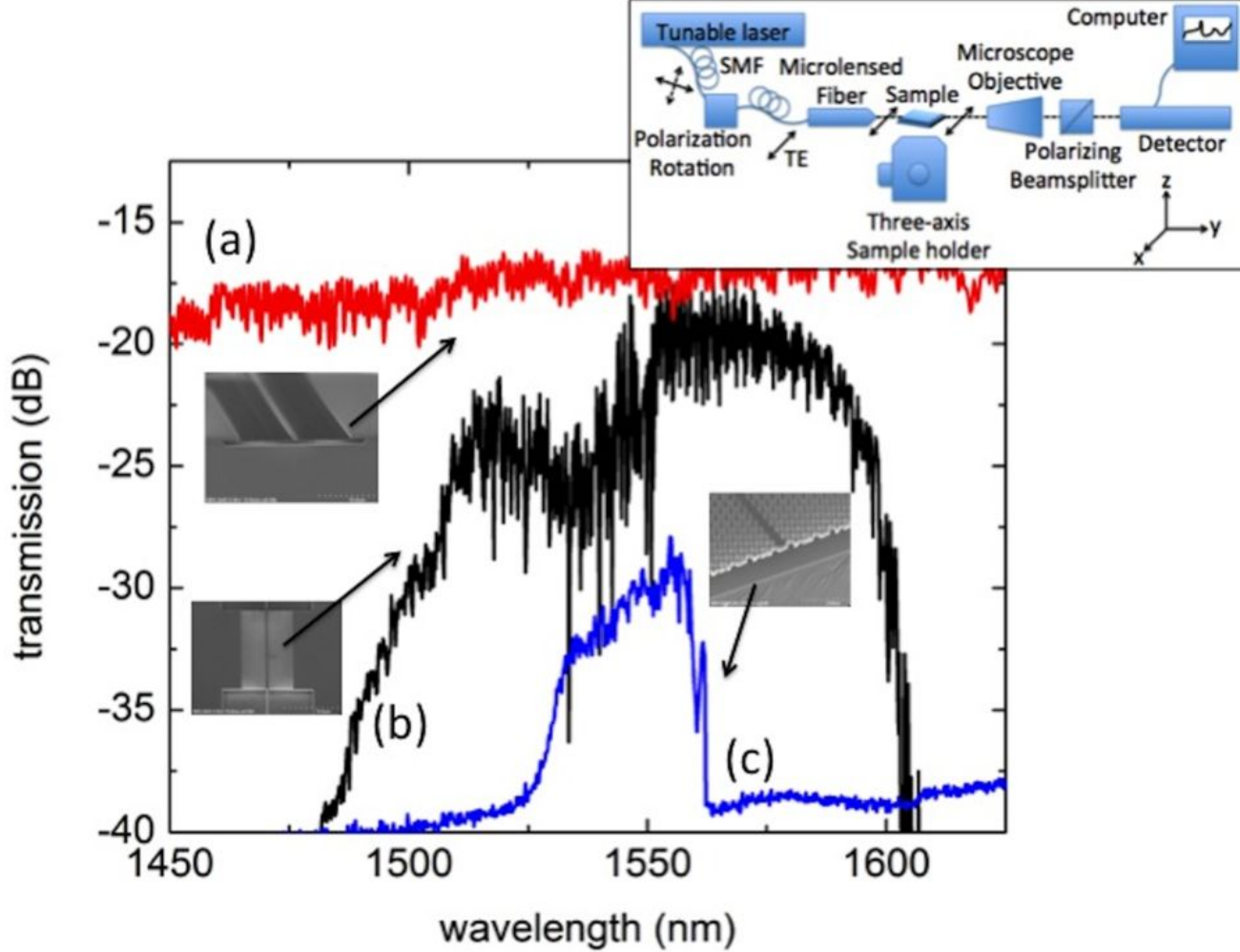